\documentclass[12pt]{article}
\newcommand{\cL}{{\cal L}}
\newcommand{\cP}{{\cal P}_L}
\newcommand{\cG}{{\cal G}_L}
\newcommand{\n}{{\bf n}}
\newcommand{\Tr}{\hbox{Tr}}
\newcommand{\tad}{\hbox{$\bigcirc\kern-3.6pt{\bf\bullet}$\kern 1.3pt}}

\begin{document}

\title{Matrix $\varphi^4$ Models on the Fuzzy Sphere and their Continuum Limits}

\author{Brian P. Dolan\footnote{
Permanent address: Dept. of Mathematical Physics, NUI, Maynooth, Ireland.}
\footnote{Email: bdolan@thphys.may.ie. }, Denjoe O'Connor\footnote{Email: denjoe@fis.cinvestav.mx}\\ 
and P. Pre\v{s}najder\footnote{
Permanent address: Dept. of Theoretical
Physics, Comenius University,\hfill\break 
${}$\hskip 18pt Mlynsk\' a Dolina, Bratislava, Slovakia.}
\footnote{Email: presnajder@fmph.uniba.sk. }\\
{\it Depto de F{\rm\'\i}sica, Cinvestav, Apartado Postal 70-543,} \\
{\it M\'exico D.F. 0730, M\'exico}\\[2ex]
}

\maketitle

\begin{abstract}{We demonstrate that the UV/IR mixing 
problems found recently for a scalar $\varphi^4$ theory 
on the fuzzy sphere are localized to tadpole diagrams and
can be overcome by a suitable modification of the action.
This modification is equivalent to normal ordering the 
$\varphi^4$ vertex.  In the limit of the commutative sphere, 
the perturbation theory of this modified action matches 
that of the commutative theory.}

\end{abstract}

\section{Introduction}
Fuzzy models have been proposed,
\cite{GrosseKlimcikPresnajder}-\cite{Bal_etal}, as a potential method
of doing ``lattice'' field theory. The basic idea is to take a
classical phase space of finite volume, quantize it and thus obtain a
space with a finite number of degrees of freedom. Things are of course
a little more complicated but essentially this idea works when the
phase space is a co-adjoint orbit, the simplest such example being the
two sphere $S^2$, with the resulting quantized space known as the
fuzzy sphere \cite{Madore}.  Field theory models on the fuzzy sphere
then possess only a finite
number of modes. The simplest field theory model of a scalar field
with $\varphi^4$ interaction was proposed in
\cite{GrosseKlimcikPresnajder} and has proved to be an excellent
testing ground for the idea of using fuzzy spaces for doing `lattice'
studies.

A potential problem with using these fuzzy spaces to get a finite
approximation to field theories and thus do `lattice' physics has
emerged due to the phenomenon of UV/IR mixing
\cite{MinwallaRaamsdonkSeiberg}.  The problem was discussed in the
Moyal plane in \cite{Kinar_etal} where the fields possess an infinite
number of modes, and consequently a regularization procedure is
needed. This phenomenon appears to be generic for field
theories in non-commutative spaces.

In a recent article Vaidya \cite{Vaidya} pointed out that this UV/IR
mixing phenomenon is present in the $\varphi^4$ model of
\cite{GrosseKlimcikPresnajder} on the fuzzy sphere.  His work was
followed by that of Chu, Madore and
Steinacker,\cite{ChuMadoreSteinacker}, who calculated explicitly the
one-loop contribution to the two point vertex function and found that
in the commutative limit, the non-planar diagram retains a residual
finite contribution over and above the expected commutative term.  The
additional term, can be seen as a nonlocal rotationally invariant
contribution to the effective action. Though this term is non-singular
for the fuzzy sphere they showed that in the planar limit
it incorporates the UV/IR mixing singularity of the Moyal plane.

The implications of this result are very serious for the program of
using the matrix model approximations to continuum field theories to
study the non-perturbative continuum behaviour. One implication is
that the scalar action considered by these authors cannot be the
correct fuzzy action for the `lattice' program outlined above.

We therefore return to the problem and study how serious it in fact is
and whether it has a natural solution. We find that indeed it has a
quite natural solution and that the problem disappears when the
interaction term in the matrix action is ``normal ordered'', i.e. when
the appropriate subtractions associated with Wick contractions of the
$\varphi^4$ term are included in the action.  The action
(\ref{normalorderedaction}) with normal ordered vertex is therefore
the correct starting point for the fuzzy lattice physics program.

The paper is organized as follows: We begin by briefly reviewing the
$*$-product, that realizes matrix multiplication at the level of
functions on the fuzzy sphere. We then review the one-loop calculation
of the two-point function, repeating the calculation using the
$*$-product. Next we discuss the 4-point vertex functions and
demonstrate that the problem anomalous contributions in that
commutative limit does not arise here. We further show, in fact, that
this problem is localized to the case of a propagator returning to the
same vertex that it leaves, i.e. that it is associated with tadpole
diagrams. We finally propose our solution for the elimination of these
unwanted contributions---normal ordering the
$\varphi^4$ vertex.  The article concludes with a discussion and
conclusions where we define the matrix model which has as limiting
theory the standard continuum model.

\section{Star products on the fuzzy sphere}

At the level of the fundamental representation the fuzzy sphere \cite{Madore}
can be defined as the orbit the 
adjoint action of $SU(2)$ on a rank one
projection operator, $\rho$, in a 2-dimensional Hilbert space, \cite{Peter} 
\cite{CPN},
with $\rho^2=\rho$,
$\rho^\dagger = \rho$ and $\Tr\rho=1$. 
At a point on the sphere parameterized by the unit vector ${\n}$ in 
${\bf R}^3$,
obtained by rotating the north pole by $g\in SU(2)$, we have
\begin{equation} 
\rho(\n)=g\rho_0g^\dagger
\end{equation}
Here $\rho_0=\left(\matrix{0&0\cr 0&1\cr}\right)$ is the projector at the north
pole, ${\n}_0=(0,0,1)$.
At the level of the $(L+1)$-dimensional representation
we construct a rank one projector $\cP(\n)$
as the $L$-fold symmetric tensor product of $\rho(\n)$.

Associated with every $(L+1)\times(L+1)$ matrix is a function on the
fuzzy sphere, defined by
\begin{equation} 
F_L(\n)=\Tr \left (\cP(\n) \hat F\right),
\label{maptofunctions}
\end{equation} 
and an associative $*$-product between two such functions is given
by 
\begin{equation} 
(F_L*G_L)(\n)=\Tr\left(\cP(\n)\hat F\hat G\right).  
\end{equation}

In terms of derivatives
\begin{eqnarray}
\label{star}
(F_L*G_L)(\n)&=&\nonumber\\
&&\kern-70pt\sum_{k=0}^L{2^k(L-k)!\over L!k!}
\Bigl(\partial_{A_1}\cdots\partial_{A_k}F_L(\n)\Bigr)
K^{A_1B_1}\cdots K^{A_kB_k}
\Bigl(\partial_{B_1}\cdots\partial_{B_k}G_L(\n)\Bigr),\nonumber\\
\end{eqnarray}
with $K^{AB}={1\over 2}(P^{AB}+iJ^{AB})$,
where $P^{AB}=\delta^{AB}-n^A n^B$ and $J^{AB}={\epsilon^{AB}}_C n^C$
so $P^2=P$ and $J^2=-P$ (indices $A,B,\ldots$ are raised and lowered with the Euclidean metric , $\delta_{AB}$, in ${\bf R}^3$).

The action of the $SU(2)$ generators on a function on the fuzzy sphere
can be written as
\begin{equation}
\cL_A F_L(\n)=\Tr\left(\cP(\n) [\hat{F},L_A]\right),
\end{equation}
where $L_A$ are the generators in the $(L+1)$-dimensional matrix
representation and 
$\cL_A=i{\epsilon_{AB}}^C n^B\partial_C$ are the generators in differential form.

The $L+1$ dimensional matrices can be expanded in terms of an
ortho-normal basis of matrices $\hat{Y}_{l,m}$ normalized so that
\begin{equation}
{4\pi\over (L+1)}\Tr\Bigl((\hat{Y}_{l',m'})^\dagger\hat{Y}_{l,m}\Bigr)=
\delta_{l'l}\delta_{m'm}.
\end{equation}
The image of these matrices under the map to functions
(\ref{maptofunctions}) we denote $Y^{L}_{l,m}({\bf n})=\Tr(\cP(\n)
\hat Y_{l,m})$. These functions are proportional to the usual
spherical harmonics $Y_{l,m}({\bf n})$ with
\begin{equation}
\label{Tnorm}
Y^L_{l,m}(\n)=T_L^{1/2}(l)\,Y_{l,m}(\n), \quad \hbox{where } \quad 
T_L(l)={L!(L+1)!\over (L-l)!(L+l+1)!}
\end{equation}
and are polynomials in $\n$ of order $l$, \cite{grklpra}.
For a general function $F_L({\bf n})$ which is the image of the matrix
$\hat{F}$ the relation (\ref{Tnorm}) takes 
the form 
\begin{equation}
F_L({\bf n})={\cal T}_L^{{1/2}}(\cL^2)F({\bf n}),
\label{F_LFreln}
\end{equation}
where the rotationally invariant operator
${\cal T}_L^{1/2}(\cL^2)$ has eigenvalues $T_L^{{1/2}}(l)$. 
Eq. (\ref{F_LFreln}) defines the relation between
functions which we denote with a subscript $L$ and those
without this subscript. The operator ${\cal T}_L(\cL^2)$ 
was introduced in \cite{Berezin}, where an alternative 
expression for it can be found.

\section{Fuzzy action}

Using the fact that 
\begin{equation}
\label{inttrace}
\int_{S^2}d^2\n \;(F_L* G_L)(\n) = {4\pi\over (L+1)} \Tr(\hat F \hat G),
\end{equation}
the integral of the $*$-product in equation (\ref{inttrace}) can be
related to the integral over an ordinary commutative product by 
decomposing the functions $F_L$ and $G_L$ into harmonics, 
(note they only contain $l\le L$), and rescaling each component by 
an $l$-dependent factor $T_L^{1/2}(l)$, then
\begin{equation}
\label{Tint}
\int d^2\n (F_L*G_L)(\n)=\int d^2\n F({\bf n}) G({\bf n}).  
\end{equation} 
where the pairs 
$F({\bf n})$, $F_L({\bf n})$ and $G({\bf n})$, $G_L({\bf n})$ are
related as in (\ref{F_LFreln}).

The na{\rm\"\i}ve matrix action for a real scalar $\varphi^4$ theory 
on $S^2_F$ at level $L$ is
\begin{eqnarray}
\label{fuzzyS}
S_L[\Phi]&=&{4\pi\over (L+1)}\Tr\left\{{1\over 2}[L_A,\Phi][L_A,\Phi]+ 
{1\over 2}r\Phi^2 +{\lambda\over 4!}\Phi^4\right\}\\
&=&\int_{S^2}d^3{\bf n}\left\{ {1\over 2}\left(\cL_A \varphi_L * \cL_A \varphi_L 
+r \varphi_L * \varphi_L \right) 
+ {\lambda\over 4!}(\varphi_L*\varphi_L*\varphi_L*\varphi_L)\right\},
\nonumber\\
\end{eqnarray}
where $\varphi_L(\n)=\Tr\left(\cP(\n)\Phi\right)$ and $\Phi=\Phi^\dagger$, so that
$\varphi_L$ is a real field. The parameters $r$ and $\lambda$ are real with 
$\lambda$ positive.

The coherent state representation of the 
free-field propagator between two points  $\n$ and $\n'$ on the fuzzy 
sphere is
\begin{equation} 
\label{fuzzyG}
\cG(\n,\n')=
\sum_{l=0}^L\sum_{m=-l}^l { {\overline Y}^L_{l,m}(\n)Y^L_{l,m}(\n')\over l(l+1)+r}.
\end{equation}
It is symmetric under interchange of its arguments, 
$\cG(\n,\n') =\cG(\n',\n)$ because $\overline Y^L_{l,m}=(-1)^m Y^L_{l,-m}$,
and can be expressed in terms of Legendre polynomials:
\begin{equation}
\label{approxfuzzyG}
\cG(\n,\n')={1\over 4\pi}\sum_{l=0}^L  
{T_L(l)(2l+1)\over l(l+1)+r}P_l(\n.\n')
\end{equation}

In the continuum limit ($L\rightarrow\infty$) 
the massless free-field propagator reads (omitting the
$l=0$ term)
\begin{equation}
\label{continuousG}
\cG'(\n,\n')={1\over 4\pi}\sum_{l=1}^\infty {(2l+1)\over l(l+1)}P_l(\n.\n')
=-{1\over 4\pi}\ln({1-\n.\n'\over2}).
\end{equation}
Note that the propagator in (\ref{fuzzyG}) is not defined with a $*$-star product, since the
two points $\n$ and $\n'$ are independent and in general distinct.

It was pointed out in \cite{Vaidya} and \cite{ChuMadoreSteinacker}
that there is a subtlety in the one-loop correction to the two-point
function for this theory.  For the commutative theory the only
infinities arise in the tadpole diagram and are accounted for by a
mass renormalization, but the fuzzy model (or fuzzy regularization)
introduces new momentum dependent terms into the effective action even
at one-loop.  These are not just wave-function renormalizations but
are new momentum dependent two-point interactions that do not
disappear in the continuum $L\rightarrow\infty$ limit.  These terms
were interpreted in \cite{Vaidya} as being due to UV/IR mixing. Their
presence however means that the fuzzy action (\ref{fuzzyS}) does not
reduce to the usual $\varphi^4$ theory in the continuum limit. It is
therefore unsatisfactory as a matrix approximation of the commutative
model.

\section{Fuzzy 2-point functions at one Loop}

The problem arises because in non-commutative $\varphi^4$ there are
two different diagrams at one loop which contribute to the quadratic
term in the effective action, a planar diagram and a non-planar
diagram, see Fig 1.

Using cyclic symmetry of the vertex it is easy to see that the 8 ways
of connecting the legs in the planar diagram contribute 
${\lambda\over 3}I_P[\varphi_L]$ 
to the effective action where
\begin{equation}
I_P[\varphi_L]=\int d^2\n
\sum_{l,m}\left\{ {\varphi_L(\n) * {\overline Y}^L_{l,m}(\n) 
* Y^L_{l,m}(\n)*\varphi_L(\n)\over l(l+1)+r}\right\},
\end{equation}
and $\sum_{l,m}:=\sum_{l=0}^L\sum_{m=-l}^l$.
Similarly the 4 ways of connecting the legs in the non-planar diagram 
contribute ${\lambda\over6}I_{NP}[\varphi_L]$ where
\begin{equation}
I_{NP}[\varphi_L]
=\int d^2\n
\sum_{l,m} \left\{{\varphi_L(\n) * {\overline Y}^L_{l,m}(\n)*\varphi_L(\n)* Y^L_{l,m}(\n)
\over l(l+1)+r}\right\},
\end{equation}
The planar diagram can be evaluated rather straightforwardly
by noting that 
\begin{equation}
\sum_{m=-l}^{l}\hat{Y}^\dag_{lm}\hat{Y}_{lm}={(2l+1)\over 4\pi}{\bf 1},
\end{equation}
so that $\sum_m{\overline Y}^L_{l,m}(\n) * Y^L_{l,m}(\n)={2l+1\over
4\pi}$ and the final $*$ can be eliminated by replacing $\varphi_L({\bf
n})={\cal T}_L^{1/2}(\cL^2)\varphi({\bf n})$ with $\varphi({\bf n})$.
So we have 
\begin{equation}
I_P[\varphi_L]=\tad(L,r)\int d^2{\bf n}{\kern 5pt} \varphi^2({\bf n}),
\end{equation}
where 
\begin{equation}
\tad(L,r)={1\over 4\pi}\sum_{l=0}^{L}{2l+1\over l(l+1)+r}
\label{commutative_tadpole}
\end{equation}
is a logarithmically divergent function in the limit of
large $L$.

The difference between the planar and non-planar diagrams is
\begin{equation}
\Delta I_2[\varphi]:=I_P[\varphi]-I_{NP}[\varphi]=
\int d^2\n \sum_{l,m} \left\{{\varphi_L(\n) * 
{\overline Y}^L_{l,m}(\n) * [Y^L_{l,m}(\n),\varphi_L(\n)]_*
\over l(l+1)+r}\right\},
\end{equation}
where $[F_L,G_L]_*:=F_L*G_L-G_L*F_L$.

Using the trace formula
(\ref{inttrace}) for the integral, and the fact that 
$({\hat Y^L_{l,m}})^\dagger =(-1)^m \hat Y^L_{l,-m}$,
cyclicity of the trace allows 
the sum and integral to be re-arranged as
\begin{equation}
\label{deltaIcom}
\Delta I_2[\varphi_L]={1\over 2}
\int d^2\n \sum_{l,m} 
\left\{{[\varphi_L(\n),{\overline Y}^L_{l,m}(\n)]_* * 
[Y^L_{l,m}(\n),\varphi_L(\n)]_*
\over l(l+1)+r}\right\}.
\end{equation}

The explicit expression for the $*$-product in (\ref{star}) shows that
the commutators start at order $1/L$ in the large $L$-limit, so one
might expect that $\Delta I_2$ would be of order $1/L^2$ and hence
vanish as $L\rightarrow\infty$.  But the problem is that the
derivatives in the $*$-product introduce factors of momenta into the
numerator, which are summed to $L$, rendering the result of order one.
Expanding $\varphi_L$ in spherical harmonics, $\varphi_L(\n)=\sum_{l=0}^L
a_{l,m}Y^L_{l,m}(\n)$, shows that there is a momentum dependent
contribution to the quadratic term in the effective action.  For
example if we restrict ourselves to fields $\varphi_L$ that are linear
functions of $\n$, i.e. fields with angular momentum $l=0$ and $l=1$
we see from (\ref{star}) that the derivatives involving $\varphi_L$
terminate at $k=1$, so we can write
\begin{equation}
[\varphi_L,{\overline Y}^L_{l,m}]_* = 
{2i\over L}(\partial_A \varphi_L) J^{AB} (\partial_B{\overline Y}^L_{l,m})=
-{2\over L}(\partial_A\varphi_L)(\cL_A{\overline Y}^L_{l,m}),
\end{equation}
and we get
\begin{equation}
\Delta I_2[\varphi_L]={2\over L^2}
\int d^2\n \sum_{l,m} 
{\{(\partial_A\varphi_L) (\cL_A{\overline Y}^L_{l,m})\}*
\{(\cL_B Y^L_{l,m})(\partial_B\varphi_L)\}
\over l(l+1)+r}.
\end{equation}
Since $\partial_A\varphi_L$ are constants, in accordance with equations 
(\ref{Tint}) and (\ref{Tnorm}) the remaining 
$*$ in this expression can be replaced with an ordinary product
provided we replace  
$Y^L_{l,m}$ with $Y_{l,m}$. Thus
\begin{equation}
\Delta I_2[\varphi_L]=-{2\over L^2}
\int d^2\n \sum_{l,m} 
{(\partial_A\varphi_L)(\cL_A{\overline Y}_{l,m})
(\cL_B Y_{l,m}) (\partial_B\varphi_L)
\over l(l+1)+r}.
\end{equation}

The problem now reduces to the evaluation of
\begin{equation}
\sum_{m=-l}^{l}\Bigl(\cL_A{\overline Y}_{l,m}(\n)\Bigr)
\Bigl(\cL_B Y_{l,m}(\n)\Bigr).
\end{equation}
Its exact form can be determined by noting that
\begin{equation}
 \sum_{m=-l}^{l}{\overline Y}_{l,m}(\n)Y_{l,m}(\n)
=\left({2l+1\over 4\pi}\right),
\end{equation}
so
\begin{equation}
\cL^2\left( \sum_{m}{\overline Y}_{l,m}Y_{l,m}\right)
=2\left\{l(l+1)\left({2l+1\over 4\pi}\right)+\sum_{m}
\Bigl(\cL_A{\overline Y}_{l,m}\Bigr)\Bigl(\cL^AY_{l,m}\Bigr)\right\}=0.
\end{equation}
from which we deduce that
\begin{equation}
\sum_{m=-l}^{l}\Bigl(\cL_A{\overline Y}_{l,m}(\n)\Bigr)
\Bigl(\cL_B Y_{l,m}(\n)\Bigr)=-{1\over 2}{(2l+1)\over 4\pi}l(l+1)P_{AB}.
\end{equation}

Finally, noting that $\partial_A\varphi_L$ only involves $l=1$ we can
write $\partial_A\varphi_L=\sqrt{L\over L+2}\partial_A\varphi$, with
$\varphi$ expanded in terms of the ordinary $Y_{lm}$ so that we obtain
\begin{equation}
\Delta I_2[\varphi_L]=\int d^2{\n}\cL_A\varphi 
{1\over L(L+2)}\sum_{l=1}^L 
\left({l(l+1)(2l+1)\over l(l+1)+r}\right)\cL_A\varphi, 
\end{equation} 
In particular we have \begin{equation} \Delta I_2[Y^L_{1,0}]={2\over
L(L+2)}\sum_{l=1}^L \left({l(l+1)(2l+1)\over l(l+1)+r}\right), \end{equation}
which is clearly finite as $L\rightarrow\infty$.  The coefficient
${2\over L(L+2)}\sum_{l=1}^L\left({l(l+1)(2l+1)\over l(l+1)+r}\right)$
was derived previously using other methods in
\cite{ChuMadoreSteinacker}.

For higher $l$ the $*$-products in the commutators in (\ref{deltaIcom}) involve terms up to $k=l$
which look as though they should be of order $1/L^{2l}$, but again the
derivatives in the numerator conspire to give a finite answer.
The full expression can be obtained by noting
that the linear operator
\begin{equation}
{\cal R}^L_{l_1}\Phi=
{4\pi\over 2l_1+1}\sum_{m_1=-l_1}^{l_1}\hat{Y}^\dag_{l_1,m_1}\Phi\hat{Y}_{l_1,m_1}
\end{equation}
is rotationally invariant and has eigenvectors $\hat{Y}_{l_2,m_2}$
and $m_2$ independent eigenvalues $\lambda^L_{l_1,l_2}$
given by
\begin{equation}
\lambda^L_{l_1,l_2}={(4\pi)^2\over (2l_1+1)(2l_2+1)(L+1)}\sum_{m_1,m_2}
\Tr [\hat{Y}^\dag_{l_1,m_1}\hat{Y}_{l_2,m_2}
\hat{Y}_{l_1,m_1}\hat{Y}^\dag_{l_2,m_2} ]
\end{equation}
Using expressions from Varshalovich et al 
\cite{VarshalovichMoskalevKhersonsky:88_book}
(equations (24) on page 46 and equation (12) page 236)
we obtain an intermediate expression in terms of Wigner 6j-symbols 
which can be summed to give 
\begin{equation}
\lambda^L_{l_1,l_2}=(L+1)(-1)^{l_1+l_2+L}
\left\{\begin{array}{ccc} l_1 & {L\over 2}&{L\over 2}\\ {} & {} & \\
l_2&{L\over 2}&{L\over 2}\end{array} \right\}.
\end{equation}
Using an explicit expression for the 6j-symbols (equation 6, page 294 
of \cite{VarshalovichMoskalevKhersonsky:88_book}) allows us to express
the eigenvalues in terms of the eigenvalues of ${\cal T}_{k}(\cL^2)$
so that we find 
\begin{equation}
\lambda^L_{l_1,l_2}=(L+1)(-1)^{l_1+l_2}\sum_{k=0}^{L}(-1)^k
\left(\begin{array}{cc}L+k+1 \\ k+1\end{array}\right)
\left(\begin{array}{cc}L \\ k \end{array}\right)T_k(l_1)T_k(l_2) .
\end{equation}
We can therefore express operator ${\cal R}_{l_1}$
in the form 
\begin{equation}
{\cal R}^L_{l_1}(\cL^2)=(L+1)(-1)^{l_1}\sum_{k=0}^{L}(-1)^{k+{\cal N}}
\left(\begin{array}{cc}L+k+1 \\ k+1\end{array}\right)
\left(\begin{array}{cc}L \\ k \end{array}\right)T_k(l_1){\cal T}_k(\cL^2) ,
\end{equation}
where ${\cal N}$ is defined by ${\cal N}Y_{l,m}(\n)=l Y_{l,m}(\n)$
and $\cL^2={\cal N}({\cal N}+1)$.
Finally we find the contribution to the effective action of the
difference of planar and non-planar diagrams can be written as
\begin{equation}
\Delta_L[\varphi_L]=\int d^2 {\bf n} \varphi_L({\bf n}){\cal Q}_L(\cL^2)\varphi_L({\bf n}),
\end{equation}
where the operator ${\cal Q}_L(\cL^2)$ is given by
\begin{equation}
{\cal Q}_L(\cL^2)={1\over 4\pi}\sum_{l=0}^{L}{2l+1\over l(l+1)+r}
\left({\cal R}^L_l(\cL^2)-1\right). 
\label{calQ_defn}
\end{equation}
The eigenvalues of this operator agree with those found in 
\cite{ChuMadoreSteinacker}.
This operator itself can clearly be expressed as the trace of a function of
$\cL^2$.

It is argued in \cite{ChuMadoreSteinacker}
that the one-loop contribution to the two-point function for external 
momentum $l$ is given, to a very good approximation, by
\begin{equation}
\Delta I_2[Y^L_{l,m}]=2\sum_{k=1}^l{1\over k}+o(1/L).
\end{equation}
We can recover this estimate by noting that $Y^L_{lm}$ are eigenfunctions
of our operator ${\cal Q}_L$ and using the expansions of \cite{ChuMadoreSteinacker}.

\section{Fuzzy 4-point functions at one loop}

In this section we argue that the problem described in the last section is
specific to the 2-point function and does not affect the 4-point function
at one loop.  The one-loop four-point function has two vertices and in the
non-commutative case these can be either planar or non-planar, giving rise to
four distinct contributions to the quartic term in the effective action.
The four distinct diagrams are given in figures 2 and 3. 
Stripping off irrelevant factors these diagrams are:
\begin{eqnarray}
\label{PP}
I_{P,P}[\varphi]:&=&\int d^2\n'd^2\n''
\sum_{l',m'}\sum_{l'',m''}\left\{
{\Bigl(\varphi_L*{\overline Y}^L_{l',m'}*Y^L_{l'',m''}*\varphi_L\Bigr)_{\n'}
\over l'(l'+1)+r}\right.\nonumber\\
&&\kern 70pt\left.\times{\Bigl(\varphi_L*{\overline Y}^L_{l'',m''}*Y^L_{l',m'}*\varphi_L\Bigr)_{\n''}
\over l''(l''+1)+r}\right\}\nonumber\\
\end{eqnarray}
\begin{eqnarray}
\label{PbarP}
I_{P,\overline{P}}[\varphi]:&=&\int d^2\n'd^2\n''
\sum_{l',m'}\sum_{l'',m''}\left\{
{\Bigl(\varphi_L*{\overline Y}^L_{l',m'}*Y^L_{l'',m''}*\varphi_L\Bigr)_{\n'}
\over l'(l'+1)+r}\right.\nonumber\\
&&\kern 70pt\left.\times{\Bigl(\varphi_L*{\overline Y}^L_{l',m'}*Y^L_{l'',m''}*\varphi_L\Bigr)_{\n''}
\over l''(l''+1)+r}\right\}\nonumber\\
\end{eqnarray}
\begin{eqnarray}
\label{NP}
I_{N,P}[\varphi]:&=&\int d^2\n'd^2\n''
\sum_{l',m'}\sum_{l'',m''}\left\{
{\Bigl(\varphi_L*{\overline Y}^L_{l',m'}*\varphi_L*Y^L_{l'',m''}\Bigr)_{\n'}
\over l'(l'+1)+r}\right.\nonumber\\
&&\kern 70pt\left.\times{\Bigl(\varphi_L*{\overline Y}^L_{l'',m''}*Y^L_{l',m'}*\varphi_L\Bigr)_{\n''}
\over l''(l''+1)+r}\right\}\nonumber\\
\end{eqnarray}
\noindent and
\begin{eqnarray}
\label{NN}
I_{N,N}[\varphi]:&=&\int d^2\n'd^2\n''
\sum_{l',m'}\sum_{l'',m''}\left\{
{\Bigl(\varphi_L*{\overline Y}^L_{l',m'}*\varphi_L*Y^L_{l'',m''}\Bigr)_{\n'}
\over l'(l'+1)+r}\right.\nonumber\\
&&\kern 70pt\left.\times{\Bigl(\varphi_L*{\overline Y}^L_{l'',m''}*\varphi_L*Y^L_{l',m'}\Bigr)_{\n''}
\over l''(l''+1)+r}\right\}.\nonumber\\
\end{eqnarray}
Surprisingly diagrams with the interchange in the second factor of
$(l',m')\leftrightarrow(l'',m'')$ (which we indicate with a bar on the
diagram label) are generally not equal to their unbarred counterpart,
even though this corresponds to a spatial twist of $\pi$ in third
dimension out of the plane of the paper.  However only the above four
are in fact distinct, the one new contribution being
$I_{P,\overline{P}}$, the others are not new but related those given by
the identities $I_{N,N}=I_{N,\overline{N}}$ and
$I_{N,P}=I_{N,\overline{P}}$.

Our aim is to show that these diagrams have no anomalous
contributions i.e. that their limiting form for $L\rightarrow\infty$
is that of the commutative theory. We begin by arguing that
the large $L$ limit of the planar contribution $I_{P,P}$ is 
the commutative one. For this we look at the contribution to
$I_{PP}$ comming from the process $\hat Y_{l_1,m_1} \hat Y_{l_2,m_2}
\rightarrow \hat Y_{l_3,m_3} \hat Y_{l_4,m_4}$
which after simplifying yields:
\begin{eqnarray}
{{\rm P\kern -6pt P}}^{l_1,m_1,l_2,m_2}_{l_3,m_3;l_4,m_4}
&=&
\sum_{l,l',l'',m}{(2l+1)(2l'+1)(2l''+1)(-1)^{l+m}\over (l'(l'+1)+r)(l''(l''+1)+r)}\nonumber\\
&&\qquad\times \left(\begin{array}{ccc} l_1 & l_2&l\\ {} & {} & \\
m_1&m_2&m\end{array} \right)
\left(\begin{array}{ccc} l_3 & l_4&l\\ {} & {} & \\
m_3&m_4&-m\end{array} \right)I^{L,l,l',l''}_{l_1,l_2,l_3,l_4}
\label{I_PP_result}
\end{eqnarray}
\vfill\eject
\noindent where we have defined
\begin{eqnarray}
I^{L,l,l',l''}_{l_1,l_2,l_3,l_4}&=&{(L+1)^2 \over (4\pi)^2}
\sqrt{(2l_1+1)(2l_2+1)(2l~_3+1)(2l_4+1)}\nonumber\\
&&\qquad\times \left\{\begin{array}{ccc} l_1 & l_2&l\\ {} & {} & \\
{L\over2}&{L\over 2}&{L\over 2}\end{array} \right\}
\left\{\begin{array}{ccc} l_3 & l_4&l\\ {} & {} & \\
{L\over2}&{L\over 2}&{L\over 2}\end{array} \right\}
\left\{\begin{array}{ccc} l' & l''&l\\ {} & {} & \\
{L\over2}&{L\over 2}&{L\over 2}\end{array} \right\}^2 .
\label{Ldependentpart}
\end{eqnarray}
Because of the triangular inequalities 
$$\vert l'-l''\vert\leq l \leq min(l_1+l_2,l_3+l_4)$$
it is first established that
\begin{equation}
\vert I_{P,\overline{P}}\vert < C\sum_{l'=0}^{L}
{(2l'+1)^2\over (l'(l'+1)+r)^2}
\end{equation}
for a positive constant $C$, depending on the external momenta.
This established that the
$L\rightarrow\infty$ limit of $I_{P,P}$ is finite and we can
therefore interchange the process of taking the large $L$ limit and
the summations.  The limiting form of
$I_{P,P}$ is then obtained by noting that for
$L\rightarrow\infty$ we have
\begin{equation}
\left\{\begin{array}{ccc} l_1 & l_2&l\\ {} & {} & \\
{L\over2}&{L\over 2}&{L\over 2}\end{array} \right\}\sim
{(-1)^{l_1+l_2+l}\over\sqrt{L+1}}
\left(\begin{array}{ccc} l_1 & l_2&l\\ {} & {} & \\
0&0&0\end{array}\right) + O(L^{-3/2})
\end{equation}
Then substituting this into (\ref{Ldependentpart}) we obtain
the leading asymptotic, which is readily seen to be the commutative
expression. 

The contribution to the diagram $I_{P,\overline{P}}$ can be obtained
from that for $I_{P,P}$ (\ref{I_PP_result}) but the corresponding
$I^{L,l,l',l''}_{l_1,l_2,l_3,l_4}$ acquires the additional phase
factor $(-1)^{l+l'+l''}$.

This procedure can be followed for the remaining diagrams, 
to establish that they also reduce to the commutative limit.
However, this is not very insightful so we shall take a different
tact. We shall show that the difference of
diagrams vanishes as $L\rightarrow\infty$, and so establish 
that they all have the same limit, which from the above we know
is the commutative limit.
\vfill\eject
Consider, therefore, difference between (\ref{PP}) and (\ref{NP}),
\begin{eqnarray} \Delta
I_4[\varphi]:&=&I_{P,P}[\varphi]-I_{N,P}[\varphi]= \int d^2\n'd^2\n''
\sum_{l',m'}\sum_{l'',m''}\nonumber\\
&&\kern-40pt\left\{{\Bigl(\varphi_L*{\overline
Y}^L_{l',m'}*[Y^L_{l'',m''},\varphi_L]_*\Bigr)_{\n'}
\Bigl(\varphi_L*{\overline
Y}^L_{l'',m''}*Y^L_{l',m'}*\varphi_L\Bigr)_{\n''} \over
\Bigl(l'(l'+1)+r\Bigr)\Bigl( l''(l''+1)+r\Bigr)}\right\}.\nonumber\\
\end{eqnarray} Again one can re-arrange traces and sums to express this as \begin{eqnarray}
\Delta I_4[\varphi]:&=& {1\over 2}\int d^2\n'd^2\n''
\sum_{l',m'}\sum_{l'',m''}\nonumber\\
&&\kern-40pt\left\{{\Bigl([\varphi_L,{\overline
Y}^L_{l',m'}]_**[Y^L_{l'',m''},\varphi_L]_*\Bigr)_{\n'}
\Bigl(\varphi_L*{\overline
Y}^L_{l'',m''}*Y^L_{l',m'}*\varphi_L\Bigr)_{\n''} \over
\Bigl(l'(l'+1)+r\Bigr)\Bigl( l''(l''+1)+r\Bigr)}\right\}.\nonumber\\
\end{eqnarray}

Na{\rm\"\i}vely one might infer from the commutators to that this
difference is of order $1/L^2$, but the analysis of the two point
function shows that one must be more careful.  In fact we shall show
that, for the four-point function, this expectation is indeed correct.
The crucial difference from the two-point function is that neither of
the propagators here closes on itself.

Expanding the commutators to first order, as was done for the 2-point
function, gives

\begin{eqnarray}
\label{fourpointdiff}
\Delta I_4[\varphi]&\approx&
{2\over L^2}\int d^2\n'd^2\n''
\left\{\Bigl(\cL_A\varphi_L(\n')\Bigr)J^{AB}(\n')\Bigl(\cL'_B\cG(\n',\n'')\Bigr)\right.\nonumber\\
&&\kern90pt\times\left.\Bigl(\cL_D\varphi_L(\n')\Bigr)J^{DC}(\n')\Bigl(\cL'_C \cG(\n',\n'')\Bigr)
\varphi^2_L(\n'')\right\},\nonumber\\
\end{eqnarray}
plus corrections down by further factors of $1/L$.

The simplest way to study the structure of $\Delta I_4[\varphi]$ as
$L\rightarrow\infty$ is to write the propagators using
(\ref{approxfuzzyG}) and observe that

\begin{equation}
\label{Gprime}
\cL^\prime_B\cG(\n'.\n'')
=i(\n'\times \n'')_B\sum_{l'=0}^L\left( {(2l'+1)P'_{l'}(\n'.\n'')(1+O(1/L^2))
\over
l'(l'+1)+r}\right),
\end{equation}
where $P'_{l'}$ is the derivative of the Legendre polynomial.  Any
singularity in the integrand in (\ref{fourpointdiff}) is coming from
terms with $l'$ and $l''$ near $L$ and most of the support of the
Legendre polynomials for such high orders is concentrated within a
region of width $1/L$ at the edges of the interval $-1\le \n'.\n''\le
1$.  This allows us to set $\n'\approx\n''$ in the argument of the
$\varphi_L$ terms in (\ref{fourpointdiff}).  Since
$P'_{l'}(1)={l'(l'+1)\over 2}$ we can approximate the sum in
(\ref{Gprime}) by ${1\over 2}\sum_{l'=0}^L\left( {(2l'+1)l'(l'+1)\over
l'(l'+1)+r}\right)\approx L^2/2$.

Finally we split the double integral up into centre-of-mass co-ordinates $\n_1$
and relative co-ordinates $\n_2$.
The relative co-ordinates are integrated over a set of measure $1/L^2$ on the fuzzy
sphere and in this region $\vert \n'\times\n''\vert \approx 1/L$.
This implies that (\ref{fourpointdiff}) is
approximately
\begin{eqnarray}
\Delta I_4[\varphi]&\approx& -{L^2\over 2}\int d^2\n_1
\Bigl(\cL_A\varphi^2_L(\n_1)\Bigr)J^{AB}(\n_1)
\Bigl(\cL_D\varphi^2_L(\n_1)\Bigr)J^{DC}(\n_1)\nonumber\\
&&\kern 50pt\times\int_{1/L^2} d^2\n_2 (\n_1\times\n_2)_B(\n_1\times\n_2)_C,
\end{eqnarray}
where the integral over $\n_2$ is over a region of area $1/L^2$ in
which $\vert \n_1\times\n_2\vert \approx 1/L$.  The second integral
above is therefore of order $1/L^4$ hence the whole expression is of
order $1/L^2$ and so vanishes in the continuum limit,
$L\rightarrow\infty$. Thus, in the $L\rightarrow\infty$ limit, there
is no difference between the planar and non-planar contributions to
the four-point coupling at one-loop.  A similar conclusion holds for
the differences (\ref{PP}) and (\ref{PbarP}), (\ref{PP}) and
(\ref{NN}), and between (\ref{NP}) and (\ref{NN}).

We have have only examined the first term in the $*$-star product
expansion, but the result can be similarly established for higher
derivatives as well, since the crucial observation was that all
derivatives of Green functions only involve Green functions with
different arguments and only the double integral ever allows these
points to co-incide.

It is relatively easy to extend the above analysis to higher vertex
functions and higher loop orders. One can, in fact, see that, for this
two dimensional theory on the fuzzy sphere, as long as the free
propagator entering the diagram does not return to the same vertex
from which it departs the diagram will limit to the commutative on and
there will be no anomalous contribution.  In other words for this
theory anomalous contributions are restricted to tadpole diagrams.

\section{Continuum $\varphi^4$-theory}

We have seen that the na{\rm\"\i}ve non-commutative action
(\ref{inttrace}) does not reproduce standard $\varphi^4$ theory in the
$L\rightarrow\infty$ limit, but instead gives a continuum theory with
a more complicated momentum dependence in the quadratic term.  However
the only source of this deviation from the standard theory is the
tadpole diagrams and this allows it to be removed in a standard
manner, by using a normal ordered vertex. By a normal ordered 
vertex we mean one that has subtracted all possible self contractions
with a suitable propagator. To this
end we define a modified bare action in matrix form, 
\begin{eqnarray}
\tilde S_L[\Phi]&=&{4\pi\over (L+1)}\Tr\left\{{1\over
2}[L_A,\Phi][L_A,\Phi] +{1\over 2} t \Phi^2 +{\lambda\over
4!}:\Phi^4:\right\},
\end{eqnarray} 
where the normal ordered interaction can be written 
\begin{eqnarray}
\Tr :\Phi^4:= \Tr\left\{ \Phi^4
-12\sum_{l,m}
\left({ \Phi {\hat Y_{lm}}^\dagger\hat Y_{lm}\Phi\over l(l+1)+t}\right)
+ 2\sum_{l,m}
\left({[\Phi,\hat Y_{lm}]^\dagger[\Phi, \hat Y_{lm}]\over l(l+1)+t}\right)
\right\}.\nonumber\\
\end{eqnarray}
The middle term is the usual tadpole subtraction that renders the two
dimensional theory finite and corresponds to normal ordering the
commutative vertex. The last term is the additional, manifestly
positive, noncommutative subtraction that is necessary to obtain the
correct commutative limit.  A more transparent form of this new action
is given by
\begin{eqnarray}
\tilde S_L[\Phi]={4\pi\over (L+1)}\Tr\left\{{1\over
2}\Phi\left(\hat{\cL}^2-{\lambda\over 2}{\cal Q}_L(\hat{\cL}^2)
+t-{\lambda\over 2}\tad(L,t)\right)\Phi +{\lambda\over
4!}\Phi^4\right\},
\label{normalorderedaction}
\end{eqnarray}
where $\hat{\cL}_i\Phi=[L_i,\Phi]$, ${\cal Q}_L$ is defined in
(\ref{calQ_defn}) and $\tad(L,t)$ in (\ref{commutative_tadpole}).
Note that the first terms can be interpreted as a momentum dependent
wavefunction renormalization since ${\cal Q}_L(\hat{\cL}^2)$ is a power
series in $\hat{\cL}^2$ which starts at order $\hat{\cL}^2$ and therefore we could
have written
\begin{equation}
\hat{\cL}^2-{\lambda\over 2}{\cal Q}_L(\hat{\cL}^2)
=\cL^2 {\cal Z}_L(\hat{\cL}^2).
\end{equation}

This theory, (\ref{normalorderedaction}), is therefore the correct
matrix model that represents a lattice regularization of commutative
theory. Note that we have further replaced the parameter $r$ with $t$,
as the two are different.  In a purely commutative theory the
replacement $r=t-{\lambda\over2}\tad(L,t)$ establishes the
relationship between the two parameters. This can be iterated to
generate all tadpole contributions.

The new quadratic terms exactly cancel all the unwanted momentum
dependent quadratic terms in the effective action arising from
non-planar diagrams in the non-commutative theory and the continuum
limit of this theory is the standard $\varphi^4$ theory in 2 dimensions.

Curiously, for this new model the direct large $L$ limit of the 
action does not give the commutative action but rather gives 
a non-local action. However this non-locality is precisely what is need
for the full quantum field theory to reproduce the commutative limit. 

For a complex scalar field on the fuzzy sphere there are two potential
vertices: $\Tr(\Phi^\dag\Phi\Phi^\dag\Phi)$ and
$\Tr(\Phi^\dag\Phi^\dag\Phi\Phi)$. The first has only a planar tadpole
and so will have no anomalous contribution, however the second has a
non-planar and therefore anomalous tadpole. Again the normal ordering
prescription removes the unwanted contribution.  In fact for any two
dimensional $\varphi^4$ model with global $O(N)$ symmetry normal
ordering the vertex will guarantee the commutative limit is recovered.

We have here only established the correct model for the two
dimensional case. A case that warrants further attention is that of
$\varphi^4$ for a four dimensional fuzzy space. The simplest example
is to take the case $S^2_F\times S^2_F$. In this case it is easily
seen from the above techniques that the problems are more severe and
that there are additional residual non-local differences for the two
and four-point functions.  This is due to the divergences that appear
in the commutative limit.  It therefore appears to be difficult to
establish precisely what fuzzy model will reproduce the commutative
limit. We leave this question open for the moment but hope to return
to it in the near future.

As a final comment we observe that the normal ordering prescription
proposed here removes the UV/IR mixing problems for the two 
dimensional $\varphi^4$ model on the Moyal plane.  This is readily 
seen since the this latter model can be recovered as a particular
scaling limit of the model considered here \cite{ChuMadoreSteinacker}.

{\bf Acknowledgments} It is a pleasure to thank Sachin Vaidya for discussions
which led to this work. We also benefited from comments especially 
of A.P. Balachandran, Oliver Jahn and Alan Stern. This work was
supported by the joint CONACyT-NSF grant E120.0462/2000 and by CONACyT 
grant 30422-E. The work of P.P. was partially supported by VEGA project
1/7069/20.

\pagebreak
\vbox{
\includegraphics{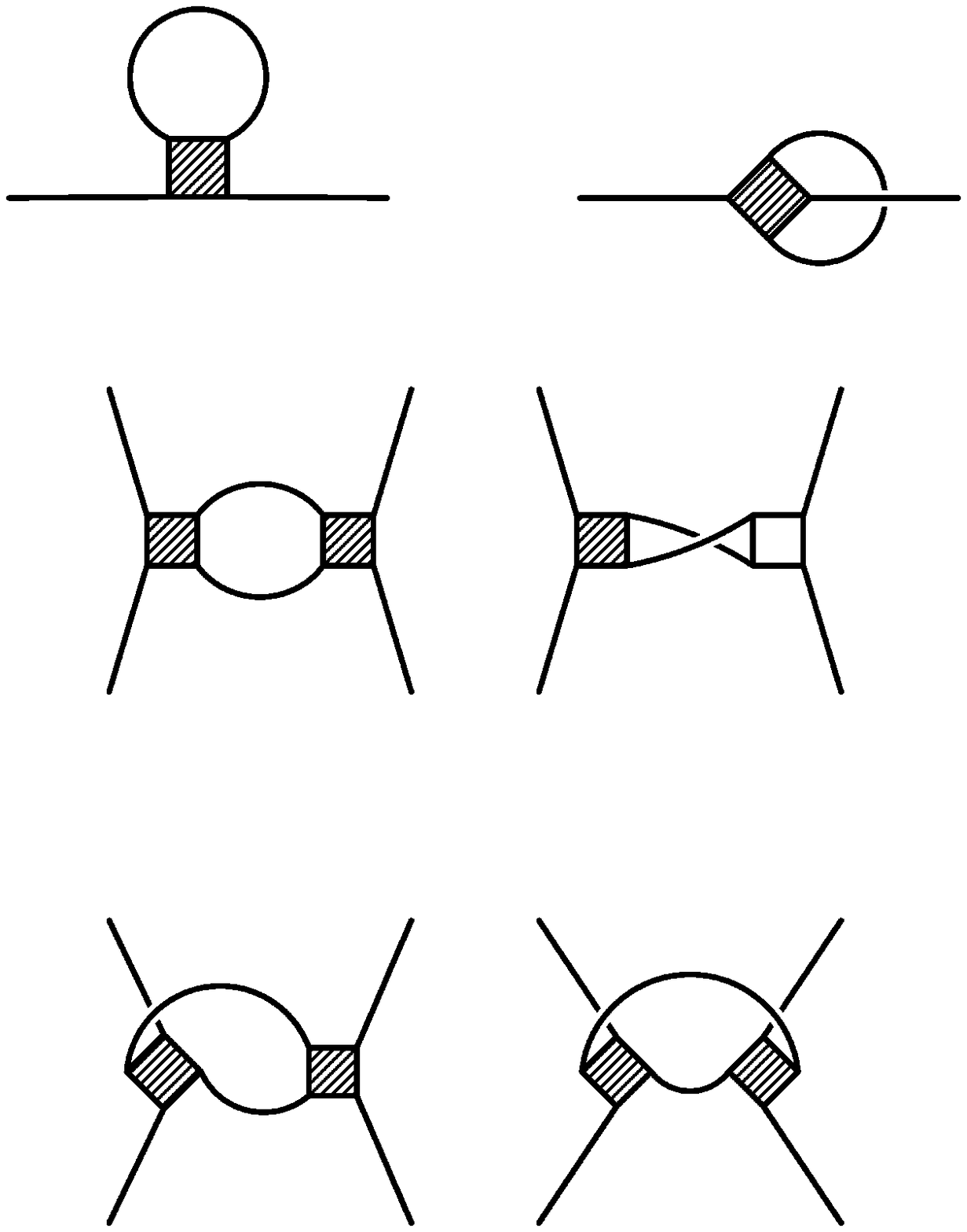}}

\vskip 3cm 

\noindent {\bf Figure 1:} One-loop diagrams contributing to the 2-point function.
The planar diagram is on the left and the non-planar on the right.

\vskip 5.5cm

\noindent {\bf Figure 2:} One-loop planar diagrams contributing to the
4-point function.  The diagram on the left has two planar vertices,
the one on the right has the right vertex rotated by $\pi$.

\vskip 5.5cm
\noindent {\bf Figure 3:} The non planar diagrams: The left one has
one planar and one non-planar vertex while the diagram on the right
has two non-planar vertices.

\end{document}